\documentstyle [a4,epsf,12pt,mystyle]{article}
\def\be{\begin{equation}}
\def\ee{\end{equation}}
\def\ba{\begin{eqnarray}}
\def\ea{\end{eqnarray}}

\def\mr{\mathrm}

\begin{document}

\heading{HYDRODYNAMIC SIMULATIONS \\ OF GALAXY FORMATION}

\smallskip
\author{Giuseppe Tormen}
{Institute of Astronomy, University of Cambridge - ENGLAND}
{and Max-Planck-Institut f\"{u}r Astrophysik, Garching - GERMANY}

\bigskip
\bigskip

\begin{abstract}{\baselineskip 0.4cm
This review is a short introduction to numerical hydrodynamics 
in a cosmological context, intended for the non specialist. 
The main processes relevant to galaxy formation are first presented. 
The fluid equations are then introduced, and their implementation 
in numerical codes by Eulerian grid based methods and by Smooth
Particle Hydrodynamics is sketched. As an application, I finally 
show some results from an SPH simulation of a galaxy cluster.}
\end{abstract}

\newpage

\section{Gas Physical Processes}

In current cosmological scenarios, the main matter components 
of the universe are some non baryonic Dark Matter (DM), which 
constitutes most of the universe, and a mixture of primordial 
gas (H, He). The DM component is decoupled from the rest of the 
universe, and interacts only through gravity. The gas component 
instead can be heated and cooled in several ways, and the physics 
involved is more complicated than in the pure gravitational case.
Fortunately, while gravity is a long range force, hydrodynamic 
processes are only important on relatively small scales, so that
on scales larger than a few Mpc the dynamics of structure formation 
can be studied with good accuracy even neglecting the gas component.
Gas dynamics, and the related radiative processes, are instead
fundamental on smaller scales, e.g. in the formation of galaxies, 
and in linking the matter distribution of the universe to the light 
distribution we actually observe$^{1)}$.
In what follows I list and discuss very briefly the main gas 
processes relevant to galaxy formation.

\subsection{Heating processes}

\begin{description}

\item [Adiabatic compression] 
is the easiest way to heat a gas. By the First Law of Thermodynamics,
compression work is converted into internal energy: 
$dQ = dU + pdV = 0 \Longrightarrow dU = - pdV$. 

\item[Viscous heating] 
is due to the small internal friction (viscosity) present in real 
gas. Velocity gradients in a gas cause an irreversible transfer
of momentum from high velocity points to small velocity ones, with
conversion of bulk velocities into random ones, i.e. into heat,
and generation of entropy. In the context of galaxy formation
viscous heating mostly occurs in shocks, which are discontinuities
in the macroscopic fluid variables due to supersonic flows. 
These arise for example during gravitational collapse, or during 
supernova (SN) explosions.

\item[Photoionization] takes place when atoms interact with the
photons of some background of soft X-rays, or UV radiation 
emitted by QSO or stars: 
e.g. $\gamma + H \rightarrow e^- + H^+$.
Observations show that at high redshift ($z \gsim 2$) hydrogen 
in the IGM is indeed ionized (Gunn-Peterson effect$^{2)}$), and 
although the origin is not clear, this is thought to be caused by 
some early generation of QSO or massive stars.
The photoionizing spectrum is usually approximated by a power law,
with flux $J(\nu) \propto (\nu / \nu_L)^{-\alpha}$, where $\nu_L$ 
is the Lyman-$\alpha$ frequency, corresponding to the hydrogen 
ionization energy, $13.6$ eV.

\end{description}

\subsection{Cooling processes}

Gas cooling is the key to galaxy formation. In fact, in our current
understanding of structure formation, the dark matter component 
of the universe first undergoes gravitational collapse, forming
dark matter halos. These provide the potential wells into which 
gas can fall and heat up by shocks, then immediately cool and form 
cold, dense, rotationally supported gas disks. In these disks 
conditions are favourable to trigger star formation, and eventually 
give rise to the galaxies we observe today$^{3)}$. The following 
cooling mechanisms are important in a cosmological context.

\begin{description}

\item [Adiabatic expansion] 
is the opposite process of adiabatic compression, with conversion
of heat into expansion work.

\item[Compton cooling]
is electron cooling against the Cosmic Microwave Background 
(CMB) through inverse Compton effect: 
$\gamma + e^-  \rightarrow \gamma + e^-$. The condition for this 
is that the temperature of the electrons is higher than the CMB
temperature: $T_e > T_\gamma$. The net energy transfer depends 
on the two densities, and on the temperature difference:
$dE/dT \propto n_e \rho_\gamma (T_e - T_\gamma)$. Since 
$E\propto n_e T_e$, the cooling time: 
$t_{cool} \equiv E/\dot{E}$ is in this case $\propto \rho_\gamma^{-1}$.
The CMB photon density decreases like $(1 + z)^4$ due to the 
expansion of the universe, therefore Compton cooling is only 
important at high redshifts (typically $z \gsim 8$), when 
the cooling timescale is smaller than the Hubble time.

\item[Radiative cooling]
is the most relevant mechanism for the cooling of primordial gas. 
It is caused by inelastic collisions between free electrons and H, 
He atoms (or their ions). Assuming that the gas is optically thin 
and in ionization equilibrium, the cooling rate per unit volume 
may be written $dE/dt \equiv \Lambda(\rho,T) = n_e n_i f(T)$, where 
$n_e$ and $n_i$ are the number densities of free electrons and of 
atoms (or ions), and $f(T)$ is called {\em cooling function}. 
The main processes are:

\begin{itemize}

\item
{\em Collisional ionization:} the inelastic scattering of a free
electron and an atom (or ion), which unbinds one electron from the
latter, e.g. $e^- + H \rightarrow H^+ + 2 e^-$.
The net cooling for the system is equal to the extraction energy.

\item
{\em Collisional excitation + line cooling:} the same situation 
as above, but the atom is only excited, and it then 
decays to the ground state, emitting a photon, e.g. 
$e^- + H \rightarrow e^- + H^* \rightarrow e^- + H + \gamma$.
This is the dominant cooling process at low 
($10^4$~K$ \lsim T \lsim 10^6~$K) temperatures.

\item
{\em Recombination:} e.g. $e^- + H^+ \rightarrow H + \gamma$. 
A free electron is captured by an ion and emits a (continuum) 
photon.

\item
{\em Bremsstrahlung:} free-free scattering between a free electron
and an ion, e.g. $e^- + H^+ \rightarrow e^- + H^+ + \gamma$. Its 
cooling rate grows with the temperature: $dE/dt \propto T^{1/2}$; 
therefore, bremsstrahlung is the dominant cooling process at high 
($T \gsim 10^6$ K) temperatures. 

\end{itemize}
\end{description}

Figure 1 shows the cooling and heating functions in different cases.

\begin{figure}[t]
\epsfxsize=\hsize\epsffile{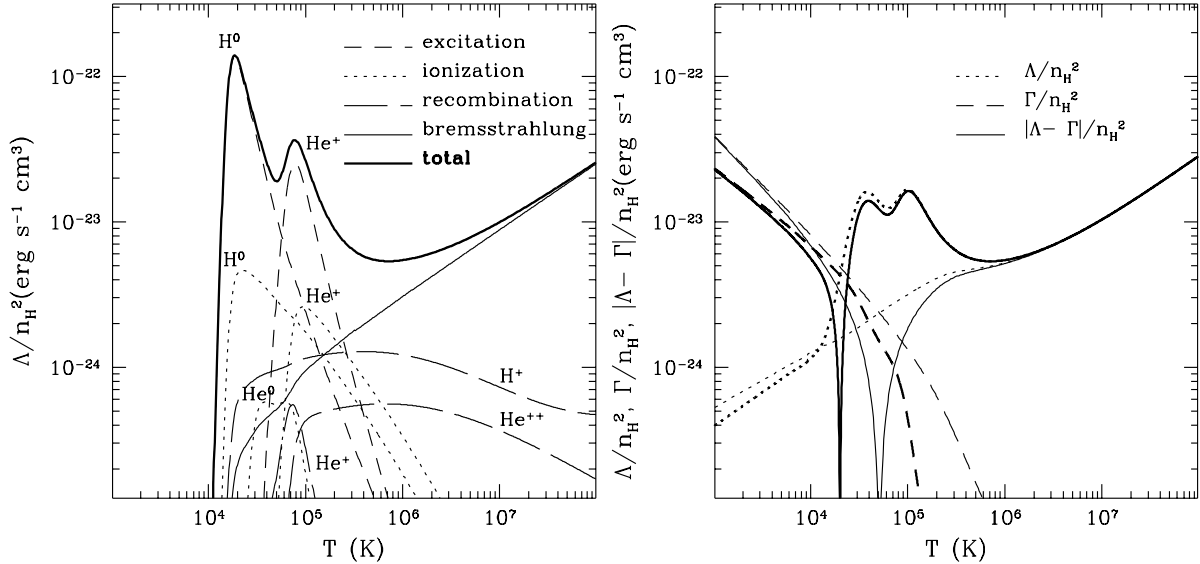}
\caption{
\baselineskip=0.4truecm
{\footnotesize
Cooling and heating functions. Left panel: cooling function 
$\Lambda(T)/n_H^2$ versus temperature, for a primordial gas. 
The different contributions to radiative cooling, and the total 
cooling curve are indicated. No photoionizing background is assumed. 
Right panel: a heating term $\Gamma/n_H^2$ is added, due to a 
photoionizing UV background with spectral index $\alpha = 1.5$. 
Its effect is both to change the ionization equilibrium (and so to 
change $\Lambda(T)/n_H^2$), and to heat the gas. Thin lines correspond 
to a gas density equal to the mean background density at redshift 
$z=5$; thick lines correspond to a density 200 times larger. 
For low density gas the effect of a UV background is dramatic, and 
line cooling is suppressed. The temperature $T_{eq}$ where 
$|\Lambda - \Gamma|$ drops to zero is called equilibrium temperature. 
At $T > T_{eq}$ the gas is cooled, at $T < T_{eq}$ it is heated. This 
figure was kindly prepared by I.Forcada using the atomic rates 
provided by T.Abel.}
}
\end{figure}

\subsection{Other processes}
Besides the heating and cooling mechanisms listed above, other 
processes may be relevant in a galaxy formation scenario. Among them:

\begin{description}

\item[Thermal conduction]
is direct transfer of heat from regions at high temperature 
to regions at lower temperature, due to the energy transport of 
diffusing electrons. The induced change in internal energy per 
unit volume is $dE/dT = \vec\nabla\left( \kappa \vec\nabla T\right)$, 
with (positive) thermal conductivity $\kappa = \kappa(T, p)$.

\item[Radiation transfer]
is important in an optically thick medium. Photons are absorbed 
by gas clouds, thermalized by multiple scattering and reemitted 
as Black Body radiation. This process depends on the optical depth 
$\tau_{opt}$ of the gas. 

\item[Star formation]
follows gas cooling as the next natural step in modeling galaxy 
formation. 
Our understanding of the detailed physics of star formation is still 
rather poor, so what is usually done is to use some empirical 
prescription to characterize the gas which is supposed to turn into 
stars. A good example of recipe$^{4)}$ is to require: {\bf\em i)} a 
convergent gas flow: $\vec\nabla\cdot\vec v <0$; {\bf\em ii)} a Jeans' 
instability criterion: the free-fall time of the gas cloud be less 
than its sound crossing time; {\bf\em iii)} a minimum number density 
of H atoms, e.g. $n_H > 0.1$ cm$^{-3}$; {\bf\em iv)} a minimum gas 
overdensity, e.g. $\rho_g \gsim \rho_V = 178 \bar{\rho_g}$, where 
$\rho_V$ is the virial density of the spherical collapse model, and 
$\bar{\rho_g}$ is the mean background gas density.
If all these conditions are satisfied, the gas will form stars at
a formation rate similar to the one observed in e.g. spiral galaxies.

Assuming some Initial Mass Function for the stars so formed, it is 
possible to compute the fraction of gas forming massive stars 
($M \geq 8 M_\odot$). These stars will explode as Type~II~SN,
each one releasing $10^{51}$~erg of energy back in the ISM, causing 
new shocks and gas heating, and triggering further star formation. 
It is also possible to include in this recipe gas release and metal 
enrichment from SN explosions. Unfortunately, at this point the 
physics of star formation is still poorly understood, and the 
resulting scenarios depend very much on the kind of assumptions 
and modeling made.

\end{description}

\section{Fluid Equations}

For our purpose it is useful to treat the gas as a continuum, 
and to resort to a hydrodynamic description. The fluid equations 
express conservation of mass (continuity equation), of momentum 
(Euler equation) and of energy. We also need an adiabatic state 
equation: $ds/dt = 0$ or $p=p(\rho,T)$. 
These equations may be written as
\be
{d\rho \over dt} = -\rho\vec\nabla\cdot\vec v, 
\label{eq:cont}
\ee
\be
\rho {d\vec v \over dt} =  -\vec\nabla p + \mr{viscosity \ terms} 
- \rho\vec\nabla\Phi,
\label{eq:eul}
\ee
\be
\rho {d\epsilon \over dt} = - p \vec\nabla\cdot\vec v 
+ \mr{viscosity \ terms} + \vec\nabla\cdot\left(\kappa\vec\nabla T\right)
+ \left( Q - \Lambda\right),
\label{eq:ene}
\ee
\be
p=p(\rho,T).
\ee

Equation~(\ref{eq:ene}) is given in terms of the specific internal 
energy $\epsilon$. In Equations~(\ref{eq:cont}) to (\ref{eq:ene})
the {\em lhs} and the first term on the {\em rhs} constitute the 
usual fluid equations for a perfect adiabatic gas. The extra terms 
on the {\em rhs} are the nonadiabatic terms introduced in the 
previous Section. Gravity is included in the Euler equation via
the gravitational potential $\Phi$. Artificial viscosity terms must
be included to enable the numerical treatment of shocks and entropy 
production, processes that do not exist in a perfect adiabatic 
gas. The terms $Q$ and $\Lambda$ denote respectively the heat sources:
photon absorption (e.g. photoionization) and energy feedback from 
SN, and the heat sinks: Compton and radiative cooling, and other 
photon emission processes. 
Including all these processes (and others, e.g. magnetic fields) 
in a recipe for galaxy formation is not an easy task. Limits arise 
both from the computational 
limitations of present day machines, and from our ignorance of the 
physics involved (e.g. the epoch and spectrum of a photoionizing 
background, or the mechanism and role of energy feedback from stars 
and SN). Moreover, some processes naturally go with others, so that 
if cooling is implemented, star formation and energy feedback should 
also be modeled. For these reasons, different physical processes may 
or may not be taken into account in different computations.
For example, current hydrodynamic simulations of structure formation 
sometimes include cooling processes, less often photoionization and
star formation. On the other hand, thermal conduction and radiative 
transport have been so far neglected. The former may play a role in 
the central part of a system, but its efficiency would depend on the 
presence of (unknown) magnetic fields. Ignoring the latter is probably
a good approximation everywhere except in very dense, optically
thick regions.
Galaxy cluster formation is a neat application of these methods, 
because cooling and star formation are less crucial here than in 
galaxy formation, so they can be ignored to first approximation. 
An example of hydrodynamic simulation of the formation of a galaxy 
cluster is briefly presented in the last Section of the paper.

\subsection{Eulerian Methods}

There are two basic ways to numerically solve the above set of 
equations: Eulerian and Lagrangian. Eulerian methods solve the 
fluid equations on a discrete grid fixed in space. In the 
traditional formulation, a Taylor expansion of the terms is 
used to build a smooth solution of the differential equations 
across different grid cells, as follows. Let $f$ be a one 
dimensional scalar field, and $F$ its flux; the 
conservation equation for $F$ is
\be
{\partial f \over \partial t} = - {\partial F \over \partial x};
\label{eq:flux}
\ee
if we Taylor expand $f(x,t)$ in time:
\be f(x, t + dt) = f(x,t) + {\partial f \over \partial t}dt
    + {1\over 2} {\partial^2 f \over \partial t^2} dt^2 
    + O(dt^3)
\label{eq:tay1}
\ee
and insert Eq.~(\ref{eq:flux}) into Eq.~(\ref{eq:tay1}), we can
substitute all time derivatives with spatial ones:
\be f(x, t + dt) = f(x,t) - {\partial F \over \partial x}dt
    + {1\over 2} {\partial \over \partial x}
      \left[{\partial F \over \partial x}{\partial F \over \partial f}\right]
      dt^2 + O(dt^3).
\label{eq:tay2}
\ee
This is the time evolution equation for $f(x,t)$, which is discretized 
and solved on a grid, to first or second order accuracy. 
Macroscopic discontinuities in the flow, like those caused by 
shocks, are mimicked by smooth solutions, obtained introducing 
explicitly, in the fluid equations, the artificial viscosity 
terms mentioned above. The underlying idea is that in a real 
fluid shocks are smooth solutions if seen at small enough scale.

More recent techniques, named {\em shock capturing schemes} or
{\em Riemann solvers}, use a different approach. They incorporate 
in the method the exact solution of a simple nonlinear problem, 
the Riemann shock tube. This solution describes the nonlinear 
waves generated by a discontinuous jump separating two constant 
states. The fluid flow is then approximated by a large number 
of constant states for which the Riemann shock tube problem is solved. 
This automatically leads to an accurate approximation of both smooth 
solutions and of shocks, with no need of explicitly introducing 
viscosity terms in the equations. This category of techniques 
includes the Piecewise Parabolic Method$^{5)}$ (PPM) and the Total 
Variation Diminishing$^{6)}$ scheme (TVD).

\subsection{Lagrangian methods: SPH}

Smooth Particle Hydrodynamics$^{7),8)}$ (SPH) is the most 
commonly used Lagrangian method in dissipative simulations of
galaxy formation. By analogy with $N$-body codes, 
where a collisionless fluid is represented with a set of discrete 
particles, SPH also uses particles to describe the evolution of a 
gas fluid.
Each particle $i$ is assigned a position $\vec r_i$, a velocity 
$\vec v_i$, a density $\rho_i$ and a specific internal energy 
$\epsilon_i$, and the fluid equations are solved at the particle's 
position, replacing the true fields with smoothed estimates, which
are obtained as local averages of the particles' properties. 
For example, for a scalar field $f$, the smoothed counterpart 
$\langle f\rangle$ is:
\be
\langle f(\vec r)\rangle = \int d^3u f(\vec u)
        W(\vec r - \vec u; h);
\label{eq:smooth}
\ee
$W(\vec r - \vec u; h)$ is the {\em smoothing kernel}, strongly
peaked at zero, so that
\be
{\mathrm{lim}}_{h \to 0} \langle f(\vec r)\rangle = \int d^3u f(\vec u)
        \delta_D(\vec r - \vec u) = f(\vec r),
\ee
with $\delta_D$ a 3 dimensional Dirac delta; $h$ is called the
{\em smoothing length}. Usually the kernel is spherically symmetric: 
$W = W(|\vec r - \vec u|; h)$, but anisotropic kernels have also 
been proposed$^{9)}$.

In practice, $\langle f(\vec r)\rangle$ is evaluated discretely at each
particle's position. Defining the particle number density at $\vec r_i$ 
as $\langle n(\vec r_i)\rangle = \rho(\vec r_i)/m_i$, in the discrete 
limit Equation~(\ref{eq:smooth}) becomes
\be
\langle f(\vec r)\rangle = 
\sum_{j=1}^N {m_j \over \rho_j} f(\vec r_j) W(|\vec r - \vec r_j|;h);
\ee
for example, for $f = \rho$ this reduces to
\be 
\langle \rho(\vec r)\rangle = 
\sum_{j=1}^N m_j W(|\vec r - \vec r_j|;h).
\ee
Typically, in SPH codes the kernel is nonzero only for e.g. 
$|\vec r - \vec u| \leq 2h$, and the smoothing length can be varied
to keep the summations to the $N = 30 - 50$ closest neighbors.

\begin{figure}[th]
\epsfxsize=\hsize\epsffile{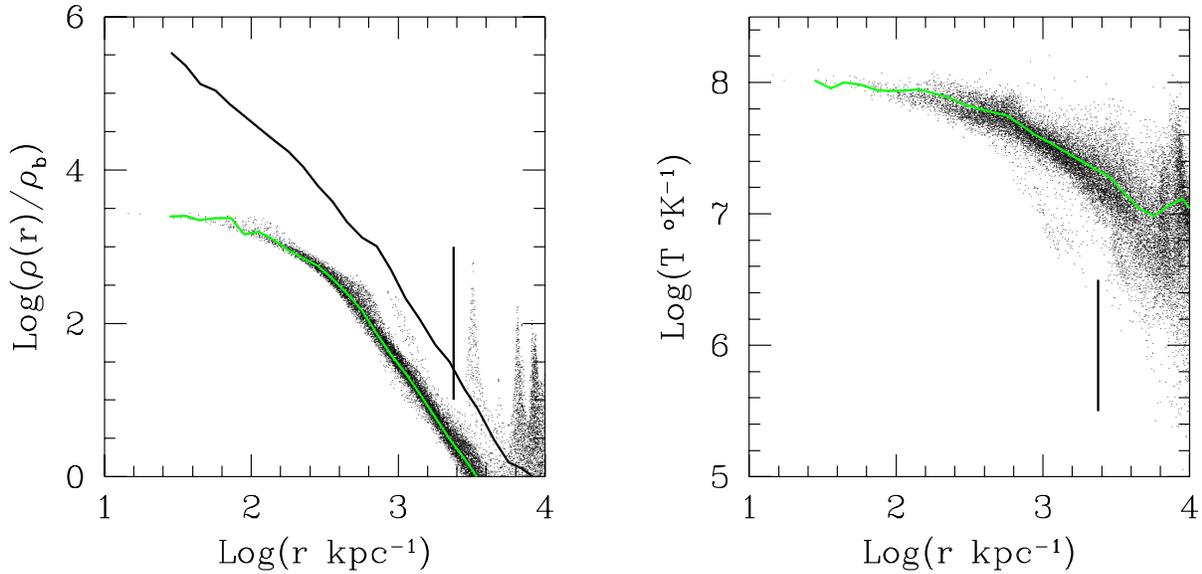}
\caption{
\baselineskip=0.4truecm
{\footnotesize
Density (left) and temperature (right) radial profiles for an SPH 
simulation of a galaxy cluster. The cluster has a virial mass of 
$7 \times 10^{14} M_\odot$ and is resolved by roughly 12000 particles 
for each species. The force resolution is $25$~kpc. More details on 
the figure are given in the main text.}
}
\end{figure}

Following this recipe, one can write the fluid equations: mass
conservation is automatically satisfied; one form of the Euler 
and energy equations is, in the adiabatic case$^{7)}$:
\be
{d \vec v_i \over dt} = - \sum_{j=1}^N m_j \left[{p_i \over \rho_i^2}
                        + {p_j \over \rho_j^2}\right]
                        \vec\nabla_i W(|\vec r_i - \vec r_j|;h);
\ee
\be
{d \epsilon_i \over dt} = {p_i \over \rho_i^2} \sum_{j=1}^N m_j
                          (\vec v_i - \vec v_j)\cdot
                          \vec\nabla_i W(|\vec r_i - \vec r_j|;h).
\ee
Nonadiabatic terms are introduced in the same way; in particular, 
shock heating is allowed by adding artificial viscosity terms.

Hybrid methods have also been proposed$^{10)}$, that make use of 
a grid to solve the fluid equations, but are Lagrangian in nature, 
since the grid itself is deformable and follows the fluid flow.
The quite different approaches of Eulerian and Lagrangian methods, 
and the variety of codes within each approach, make it difficult 
to compare results coming from different codes, and to interpret 
the comparison. An attempt has however been done$^{11)}$ by 
evolving, from some high redshift to the present time, the same 
initial conditions of a cosmological model, using three different 
Eulerian codes and two different SPH codes; the results seem to
indicate that Eulerian codes have better resolution on large scales
and low $\rho$, low $T$ regions, while SPH codes can better
resolve small scales and high $\rho$, high $T$ regions.

\section{Applications: Formation of a galaxy cluster}

Some results from an SPH simulation of a galaxy cluster are
shown in Figure~2, as an example of application of the theory
summarized in the previous Sections. The gas processes modeled 
are adiabatic heating and cooling, and viscous heating, while other 
cooling processes, photoionization and star formation were 
neglected, since they are not crucial to this particular problem.

The simulation has an Einstein-de Sitter background universe, 
with $H_0 = 50$ km s$^{-1}$ Mpc$^{-1}$, and with scale free 
power spectrum of perturbation $P(k) \propto k^{-1}$. 
The left panel shows the gas density (points), and the DM and gas 
mean density profiles (black and grey curve respectively), as a 
function of the distance $r$ from the cluster center; 
the vertical bar indicates the virial radius of the cluster. 
Note how the gas is less centrally concentrated than the DM. 
The density spikes in the gas profile are infalling substructure. 
The right panel shows the gas temperature and the mean temperature 
profile at different radii: the cluster is not isothermal, and its 
temperature drops by a factor of five from the center to the virial 
radius. During collapse the gas is shock heated to about $10^7$~-~$10^8$~K
as seen both in the main system and in the infalling substructures.

\subsection*{Acknowledgments}
I would like to thank Bruno Guiderdoni for inviting me to give this
talk. Thanks also to Bhuvnesh Jain, Ignasi Forcada and Simon White 
for comments on the manuscript. Financial support from an EC-HCM
fellowship is gratefully acknowledged.

\end{document}